\documentclass[aip,reprint]{revtex4-1}
\usepackage{hyperref}
\usepackage{graphicx}

\draft 

\begin{document}


\title{Highly textured oxypnictide superconducting thin films on metal substrates} 

\author{Kazumasa\,Iida}
\email[Electronical address:\,]{iida@nuap.nagoya-u.ac.jp}
\altaffiliation[Present address:\,]{Department of Crystalline Materials Science, Nagoya University, Chikusa, Nagoya 464-8603, Japan}
\author{Fritz\,Kurth}
\affiliation{Institute for Metallic Materials, IFW Dresden, D-01171 Dresden, Germany}
\author{Masashi\,Chihara}
\author{Naoki\,Sumiya}
\affiliation{Department of Crystalline Materials Science, Nagoya University, Chikusa, Nagoya 464-8603, Japan}
\author{Vadim\,Grinenko}
\affiliation{Institute for Metallic Materials, IFW Dresden, D-01171 Dresden, Germany}
\author{Ataru\,Ichinose}
\author{Ichiro\,Tsukada}
\affiliation{Central Research Institute of Electric Power Industry, 2-6-1 Nagasaka, Yokosuka, Kanagawa 240-0196, Japan}
\author{Jens\,H{\"a}nisch}
\affiliation{Institute for Metallic Materials, IFW Dresden, D-01171 Dresden, Germany}
\author{Vladimir\,Matias}
\affiliation{iBeam Materials, Inc., 2778A Agua Fria Street, Santa Fe, NM 87507, U.S.A.}
\author{Takafumi\,Hatano}
\affiliation{Department of Crystalline Materials Science, Nagoya University, Chikusa, Nagoya 464-8603, Japan}
\author{Bernhard\,Holzapfel}
\affiliation{Karlsruhe Institute of Technology, Institute for Technical Physics, Hermann von Helmholtz-Platz 1, D-76344 Eggenstein-Leopoldshafen, Germany}
\author{Hiroshi\,Ikuta}
\affiliation{Department of Crystalline Materials Science, Nagoya University, Chikusa, Nagoya 464-8603, Japan}



\date{\today}

\begin{abstract}
Highly textured NdFeAs(O,F) thin films have been grown on ion beam assisted deposition (IBAD)-MgO/Y$_2$O$_3$/Hastelloy substrates by molecular beam epitaxy. The oxypnictide coated conductors showed a superconducting transition temperature ($T_{\rm c}$) of 43\,K with a self-field critical current density ($J_{\rm c}$) of $7.0\times10^4\,{\rm A/cm^2}$ at 5\,K, more than 20 times higher than powder-in-tube processed SmFeAs(O,F) wires. Albeit higher $T_{\rm c}$ as well as better crystalline quality than Co-doped BaFe$_2$As$_2$ coated conductors, in-field $J_{\rm c}$ of NdFeAs(O,F) was lower than that of Co-doped BaFe$_2$As$_2$. These results suggest that grain boundaries in oxypnictides reduce $J_{\rm c}$ significantly compared to that in Co-doped BaFe$_2$As$_2$ and, hence biaxial texture is necessary for high $J_{\rm c}$.      
\end{abstract}


\maketitle
Electron or hole doped $AE$Fe$_2$As$_2$ ($AE=$Ba and Sr) and Fe(Se,Te) show high upper critical fields with low anisotropies at low temperatures that offer a unique possibility for high field magnet applications\cite{01,02}. Thanks to the recent progress of processing technology, powder-in-tube (PIT) processed $AE$Fe$_2$As$_2$ yields high performance of in-field critical current density ($J_{\rm c}$)\cite{03,04,05}. Furthermore, Co-doped BaFe$_2$As$_2$ and Fe(Se,Te) on ion beam assisted deposition (IBAD)-MgO/Y$_2$O$_3$/Hastelloy with comparable $J_{\rm c}$ to those on single crystalline substrates have been demonstrated, although the texture quality of the formers is inferior to the latters\cite{15,06,07,08}. These results indicate that $J_{\rm c}$ properties are quite robust against grain boundaries (GBs) in those materials. Indeed, the critical angle for Co-doped BaFe$_2$As$_2$ at which $J_{\rm c}$ starts to being degraded is twice as large as cuprates\cite{09,10}.

On the other hand, $Ln$FeAs(O,F) ($Ln=$Sm, Nd) is expected to have potentially higher $J_{\rm c}$ than other Fe-based superconductors due to the highest $T_{\rm c}$ around 55\,K\cite{13,Uemura} among the Fe-based superconductors except for monolayer FeSe\cite{14}. However, the level of $J_{\rm c}$ in PIT-processed SmFeAs(O,F) wire is around $4.0\times10^3\,{\rm A/cm^2}$ even in low fields at 4.2\,K\cite{12}, much lower than $AE$Fe$_2$As$_2$ wires\cite{03,04,05}. A fundamental reason for the low $J_{\rm c}$ in oxypnictide may be due to higher sensitivity against the grain misorientation than other Fe-based superconductors, since $Ln$FeAs(O,F) ($Ln=$Sm, Nd) has more anisotropic crystal structure and superconducting properties, which resembles high-$T_{\rm c}$ cuprates. Those facts imply that biaxial textured forms are necessary for high $J_{\rm c}$, indicating that the 2nd generation YBa$_2$Cu$_3$O$_7$ coated conductor technology might be suitable for realizing oxypnictide superconducting wires. However, no studies on oxypnictide coated conductors have been reported to date due presumably to the difficulties in fabricating epitaxial $Ln$FeAs(O,F) thin films on technical substrates.

\begin{figure}
	\centering
		\includegraphics[width=7cm]{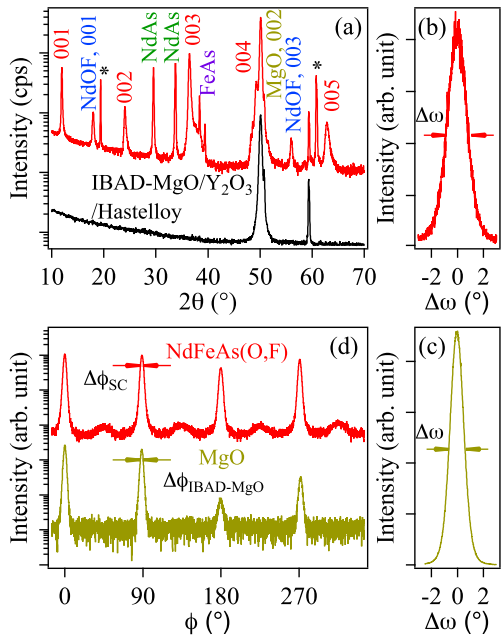}
		\caption{(Color online) Summary of structural characterization by x-ray diffraction: (a) the $\theta\rm/2\theta$\,-\,scans in Bragg-Brentano geometry using Co-K$_\alpha$ radiation. The red and black traces are NdFeAs(O,F) coated conductor and IBAD-MgO/Y$_2$O$_3$/Hastelloy substrate, respectively. Unindentified phase is marked by a star. (b) The 005 rocking curve of NdFeAs(O,F) using Cu-K$_\alpha$ radiation. (c) The 002 rocking curve of MgO. (d) The 102 $\phi$-scans of NdFeAs(O,F) and the 220 $\phi$-scans of MgO measured in a texture goniometer operating with Cu-K$_\alpha$ radiation.} 
\label{fig:figure1}
\end{figure}

Here we report on a preparation of NdFeAs(O,F) thin films on MgO-templated Hastelloy by means of molecular beam epitaxy (MBE). By comparing with similar studies on Co-doped BaFe$_2$As$_2$ coated conductors\cite{15}, we point out the necessity of biaxial texture in oxypnictide wires and tapes for high $J_{\rm c}$.

The NdFeAs(O,F) thin films were prepared on IBAD-MgO/Y$_2$O$_3$/Hastelloy by MBE. The IBAD-MgO/Y$_2$O$_3$/Hastelloy has been provided by iBeam Materials, Inc\cite{16}. A mother compound of NdFeAsO was deposited on MgO templated Hastelloy, followed by the deposition of a NdOF cap layer for F diffusion to the NdFeAsO layer. The detailed fabrication can be found in Ref.\,\onlinecite{17}. The respective thickness of NdFeAs(O,F) superconducting layer, MgO template and Y$_2$O$_3$ buffer layer were approximately 90\,nm, 180\,nm and 10\,nm confirmed by cross-sectional transmission electron microscope (TEM), which will be shown later.

Phase purity and texture quality of the films were examined by x-ray diffraction (Fig.\,\ref{fig:figure1}). The $\theta\rm/2\theta$\,-\,scan, fig.\,\ref{fig:figure1}\,(a), shows  00$l$ reflections of NdFeAs(O,F) and NdOF together with the 002 reflection of MgO, indicative of $c$-axis texturing. Albeit the film was $c$-axis oriented, NdAs, FeAs and unidentified impurity phase were detected. The $\omega$\,-\,scan for the 005 reflection of NdFeAs(O,F) in fig.\,\ref{fig:figure1}\,(b) shows a full width at half maximum (FWHM, $\Delta\omega$) of 1.72$^{\circ}$, which is a similar value to the textured MgO template[($\Delta\omega=1.24^{\circ}$), fig.\,\ref{fig:figure1}\,(c)]. The 102 $\phi$-scans of NdFeAs(O,F) reveal a small amount of 45$^{\circ}$ in-plane rotated grains, as shown in fig.\,\ref{fig:figure1}\,(d). However, the main reflections were observed at every 90$^{\circ}$, indicative of highly biaxial textured NdFeAs(O,F) films. Here, the epitaxial relation is identified as (001)[100]NdFeAs(O,F)$\|$(001)[100]MgO. The average $\Delta\phi$ of NdFeAs(O,F) and MgO are 3.38$^{\circ}$ and 2.88$^{\circ}$, respectively. Hence $\Delta\omega$ and $\Delta\phi$ of NdFeAs(O,F) are similar to those of the underlying MgO template, indicating that the texture is well transferred to NdFeAs(O,F). We have also confirmed that the NdOF cap layer is biaxially textured and its epitaxial relation is (001)[110]NdOF$\|$(001)[100]NdFeAs(O,F), which is an observation similar to SmFeAs(O,F) thin films\cite{18}.
  
\begin{figure}
	\centering
		\includegraphics[width=\columnwidth]{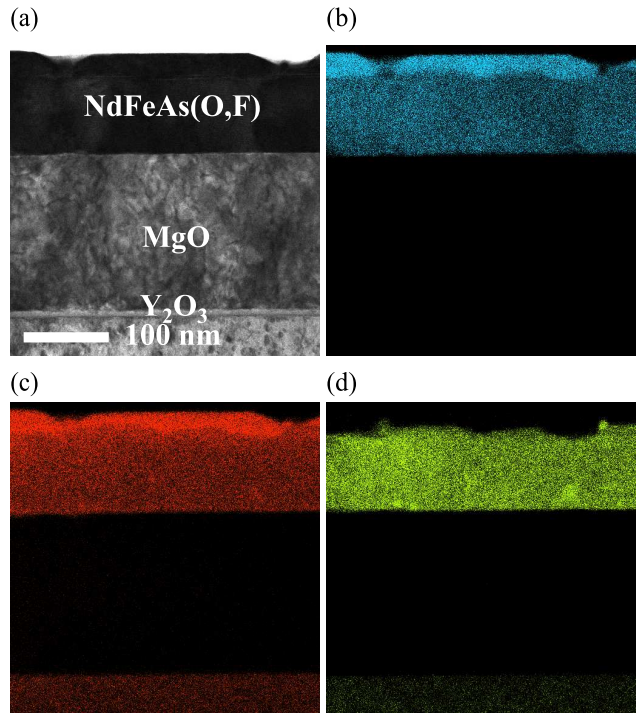}
		\caption{(Color online) (a) Cross-sectional scanning TEM image of the NdFeAs(O,F) thin film. The respective thickness of NdFeAs(O,F) superconducting layer, MgO template and Y$_2$O$_3$ buffer layer are approximately 90\,nm, 180\,nm and 10\,nm. (b) Elemental Nd, (c) F, and (d) Fe mappings measured by energy dispersive x-ray spectroscopy.} 
\label{fig:figure2}
\end{figure}

\begin{figure}
	\centering
		\includegraphics[width=\columnwidth]{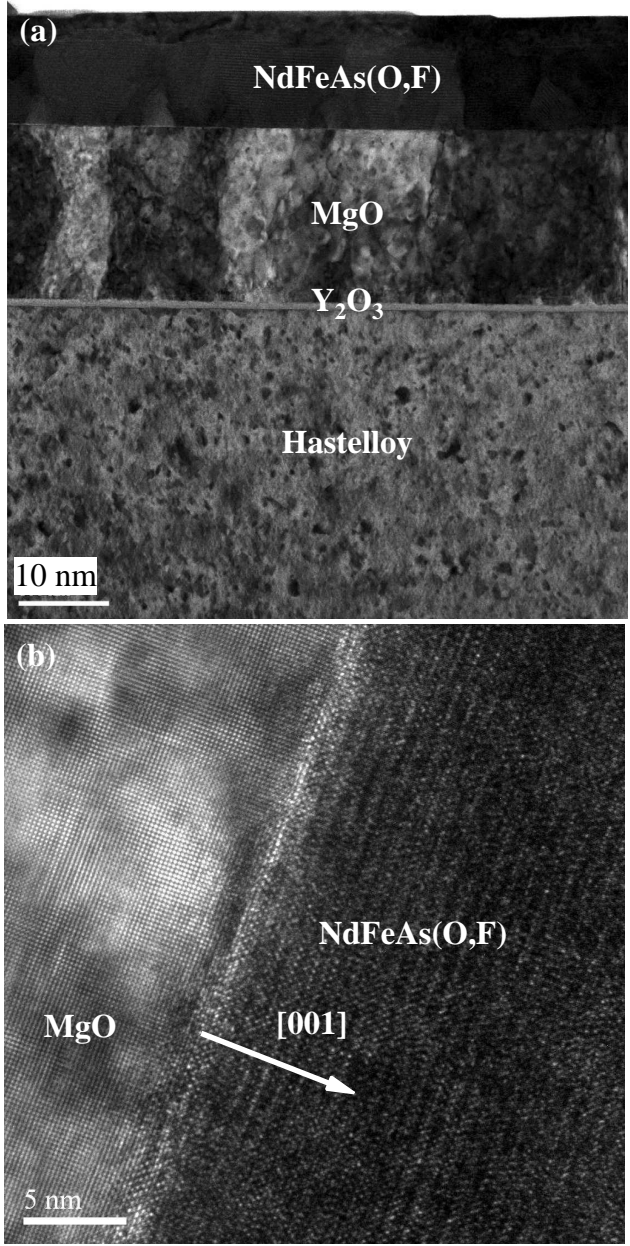}
		\caption{(a) Cross-sectional scanning TEM image of the NdFeAs(O,F) thin film, which is an area different from fig.\,\ref{fig:figure2}(a). (b) High-resolution TEM image of the NdFeAs(O,F) thin film in the vicinity of the MgO textured template / NdFeAs(O,F) film interface.} 
\label{fig:figure3}
\end{figure}

In order to check the homogeneity of the NdFeAs(O,F), microstructural investigation by TEM together with elemental mappings were carried out (fig.\,\ref{fig:figure2}). In the superconducting layer, some areas have different contrast, which originates from misoriented grains transferred from the underlying MgO templates. As can be seen in fig.\,\ref{fig:figure3}(a), the MgO layer with white and black in-contrast yields different types of NdFeAs(O,F) grains (i.e., texture transferring from the buffer layer). It is further obvious from fig.\,\ref{fig:figure2}(a) that trapezoid shaped NdOF layer was grown on NdFeAs(O,F), which is proved by the elemental mappings shown in fig.\,\ref{fig:figure2}(b) and (c) [i.e., high concentration of both Nd and F] and x-ray $\theta\rm/2\theta$\,-\,scan shown in fig.\,\ref{fig:figure1}\,(a). Whilst Nd is homogeneous in the superconducting layer, Fe is segregated in the vicinity of the interface [fig.\,\ref{fig:figure2}\,(d)]. Although it is less clear, we note that the area where Fe is segregated seems to be also rich in F.   

Shown in fig.\,\ref{fig:figure3}(b) is the high-resolution TEM image of the NdFeAs(O,F) thin film in the vicinity of the MgO textured template / NdFeAs(O,F) film interface. Clearly a relatively sharp interface is observed between NdFeAs(O,F) and MgO, which is similar to the Co-doped Ba-122 and Fe(Se,Te) thin films on MgO single crystalline substrates prepared by pulsed laser deposition\cite{19,20}. These results indicate that MgO single crystalline substrates or textured MgO templates are suitable for epitaxial growth of Fe-based superconducting thin films, similarly to CaF$_2$ substrate\cite{Tsukada,21,Uemura,Ichinose}.

\begin{figure}
	\centering
			\includegraphics[width=7cm]{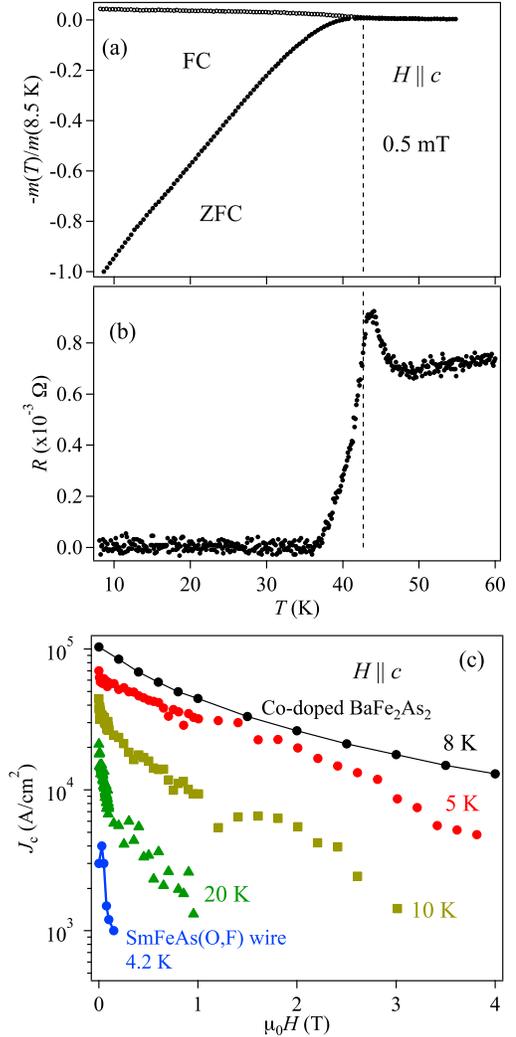}
		\caption{(Color online) (a) The normalized magnetization curves $-m(T)/m(8.5\,{\rm K})$ of the NdFeAs(O,F) coated conductor for zero field-cooled (ZFC) and field-cooled (FC) branches measured by a SQUID magnetometer. A small magnetic field of 0.5\,mT was applied parallel to the $c$-axis of the film. The data were normalized to the ZFC value at 8.5\,K. (b) The transport measurements for the same film conducted in a PPMS by a four-probe method. (c) The $J_{\rm c}-H$ properties of NdFeAs(O,F) for $H\parallel c$ measured in a SQUID magnetometer. In-field transport $J_{\rm c}$ of Co-doped BaFe$_2$As$_2$ coated conductor and PIT-processed SmFeAs(O,F) wire are also plotted in the same figure for comparison\cite{12,15}.}
\label{fig:figure4}
\end{figure}

The NdFeAs(O,F) film exhibits a $T_{\rm c}$ of 43\,K at which spliting between zero field-cooled (ZFC) and field-cooled (FC) branches of the normalized magnetization curves $-m(T)/m(8.5\,{\rm K})$ is observed as shown in fig.\,\ref{fig:figure4}(a). Here magnetic measurements have been carried out by means of a superconducting quantum interference device (SQUID, Quantum Design) magnetometer and the data were normalized to the ZFC value at 8.5\,K. On the other hand, zero resistance was observed at 37\,K, measured in a Physical Property Measurement System (PPMS, Quantum Design) by a four-probe method [fig.\,\ref{fig:figure4}(b)]. This difference is due to a large bias current of 100\,mA, leading to a shift of zero resistance temperature to lower temperatures. A sudden increase in the resistance before the superconducting transition is a general observation for films on conducting substrates, where the resistivity of the films in the normal state is higher than that of the substrates\cite{22}.

Shown in fig.\,\ref{fig:figure4}(c) is in-field $J_{\rm c}$ property of the NdFeAs(O,F) coated conductor measured by a magnetization method using a SQUID magnetometer. Magnetic $J_{\rm c}$ was evaluated by the following formula:  $J_{\rm c}=\frac{20\Delta m}{(a-a^2/3b)V}$, where $\Delta m$ is the difference between the magnetic moments observed in the magnetic hysteresis loops for increasing and decreasing field cycles, $a$ and $b$ are the film dimensions for the $ab$-plane, and $V$ is the superconducting volume\cite{Bean}. As can be seen in fig.\,\ref{fig:figure4}(c), the self-field $J_{\rm c}$ of NdFeAs(O,F) coated conductor reaches $7.0\times10^4\,{\rm A/cm^2}$ at 5\,K, which is more than 20 times higher than PIT-processed SmFeAs(O,F)\cite{12}. Additionally, the oxypnictide coated conductor shows better in-field performance than PIT-processed SmFeAs(O,F). However, $J_{\rm c}$ of NdFeAs(O,F) film on MgO single crystalline substrate exceeds $10^4{\rm A/cm^2}$ even at 35\,T ($H\parallel c$) at 4.2\,K\cite{Chiara}, indicating that there is much room for optimization. Compared to the Co-doped BaFe$_2$As$_2$ coated conductor\cite{15}, the level of $J_{\rm c}$ for NdFeAs(O,F) is relatively low. Note that the zero resistance temperature of Co-doped BaFe$_2$As$_2$ is 17.5\,K, much lower than that of NdFeAs(O,F). Furthermore, in-field $J_{\rm c}$ for Co-doped BaFe$_2$As$_2$ shown in fig.\,\ref{fig:figure4}(c) was measured at 8\,K. By considering the same reduced temperature (i.e., $t=T/T_{\rm c,0}=8/17.5=0.457$ for Co-doped BaFe$_2$As$_2$), one can expect that $J_{\rm c}$ of NdFeAs(O,F) at 16.9\,K, which is supposed to be located between the curves measured at 10 and 20\,K, is reduced significantly by only small magnetic fields. It is also worth mentioning that the $\Delta\phi$ value of NdFeAs(O,F) is smaller than that of Co-doped BaFe$_2$As$_2$\cite{15}, as shown in Table~\ref{tab:table1}. These results indicate that NdFeAs(O,F) shows weak-link behavior, which necessitates biaxial texture for high $J_{\rm c}$. Nevertheless bicrystal experiments on $Ln$FeAs(O,F) are necessary to quantitatively characterize the GBs properties.

\begin{table}
\caption{\label{tab:table1}The average FWHM values ($\Delta\phi_{\rm SC}$) of NdFeAs(O,F), Co-doped BaFe$_2$As$_2$ on IBAD-MgO/Y$_2$O$_3$/Hastelloy and IBAD-MgO. The data for Co-doped BaFe$_2$As$_2$ were taken from Ref.\onlinecite{15}. The respective reflections for NdFeAs(O,F), Co-doped BaFe$_2$As$_2$ and MgO are 102, 103 and 220.}
\begin{ruledtabular}
\begin{tabular}{lcr}
Films&$\Delta\phi_{\rm SC}$&$\Delta\phi_{\rm IBAD-MgO}$ \\
\hline
NdFeAs(O,F) & $3.38^\circ$ &  $2.88^\circ$ \\
Co-doped BaFe$_2$As$_2$ & $5.13^\circ$ & $5.98^\circ$ \\
\end{tabular}
\end{ruledtabular}
\end{table}
 
In summary, textured NdFeAs(O,F) thin films have been realized on IBAD-MgO/Y$_2$O$_3$/Hastelloy substrates by MBE. The oxypnictide coated conductors showed a $T_{\rm c}$ of 43\,K with a self-field $J_{\rm c}$ of $7.0\times10^4\,{\rm A/cm^2}$ at 5\,K, more than 20 times higher than PIT processed SmFeAs(O,F) wires. However, the current-limiting effects by GBs in $Ln$FeAs(O,F) seem to be more serious than in $AE$Fe$_2$As$_2$ and Fe(Se,Te). Hence it is necessary to align grains biaxially in oxypnictide wires and tapes for high $J_{\rm c}$.
 
\begin{acknowledgments}
The authors thank M.\,K\"{u}hnel and U.\,Besold for their technical support. The research leading to these results has received funding from European Union's Seventh Framework Programme (FP7/2007-2013) under grant agreement number 283141 (IRON-SEA). This research has been also supported by Strategic International Collaborative Research Program (SICORP), Japan Science and Technology Agency.
\end{acknowledgments}


%


%

\end{document}